\documentclass[12pt]{article}
\usepackage[dvips]{graphicx}

\def\lromn#1{\uppercase\expandafter{\romannumeral#1}}

\def\Slash#1{{\ooalign{\hfil$#1$\hfil\crcr\hfil$/$\hfil}}}

\begin{document}
  
\begin{titlepage}

\begin{center}

\hfill KEK-TH-1120 \\
\hfill \today

\vspace{1cm}
{\large\bf Hunting for the Top Partner in the Littlest Higgs Model with
T-parity at the LHC}
\vspace{1.5cm}

{\bf Shigeki Matsumoto}\footnote{E-mail: smatsu@post.kek.jp},
{\bf Mihoko M Nojiri}\footnote{E-mail: nojiri@post.kek.jp},
and
{\bf Daisuke Nomura}\footnote{E-mail: dnomura@post.kek.jp}
\vskip 0.2in
{\it Theory Group, KEK, 1-1 Oho, Tsukuba, 305-0801, Japan}

\vskip 1in

\abstract{
We study the Littlest Higgs model with T-parity at the LHC through pair
productions of the T-odd top quark partner ($T_-$) which decays into the top
quark and the lightest T-odd particle. We identify the region of parameters
favored by the electroweak and cosmological considerations. The signal and
background events are simulated with fast detector simulation to study the
discovery potential at the LHC. We find that the hemisphere analysis recently
proposed by the CMS collaboration is very useful to separate the signal 
from the
$t\bar{t}$ background. We discuss the observability of the top tagged signal in
the effective mass ($M_{\rm eff}$) versus the transverse missing energy
($E_{\rm Tmiss}$) plane. We show that, for all our sample parameter sets with
$M_{T_-}\leq$ 900~GeV, the excess of the signal over the background can be
visible as a bump structure in the $E_{\rm Tmiss}$ distribution for 50 fb$^{-1}$
at relatively high $M_{\rm eff}$ intervals.}

\end{center}
\end{titlepage}
\setcounter{footnote}{0}

\section{Introduction}

The Higgs sector in the Standard Model (SM) receives large quadratic mass
corrections from top and gauge boson loop diagrams. New symmetries involving
top-Higgs and gauge-Higgs sectors below ${\cal O}$(1 TeV) are proposed to
remove the corrections. One of the physics targets of the ATLAS \cite{ATLASTDR}
and CMS \cite{CMSTDR} experiments at the LHC is to find new particles predicted
by such symmetries.

The most important new physics candidate is the Minimal Supersymmetric Standard
Model (MSSM)\cite{MSSM}. This model predicts quark and lepton partners with
spin 0 (squark and slepton) and those of gauge and Higgs bosons with spin 1/2
(gauginos and Higgsinos). Thanks to the cancellation among boson and fermion
loop diagrams, the quadratic corrections to the Higgs mass term completely
vanish. At the LHC, strongly interacting supersymmetric particles such as
squark and gluino will be copiously produced. They decay into relatively light
electroweak (EW) superparters, such as the chargino, neutralino, and slepton.
The lightest supersymmetric particle (LSP) is stable due to the R-parity of the
model. The decay products of the squark and gluino contain at least one
LSP.  The LSP escapes
from detector without any energy deposit, giving a missing momentum signature
to the events. The signature of supersymmetric particles has been studied
intensively by many groups \cite{ATLASTDR,CMSTDR,SUSYLHC}.

Alternative scenarios which do not rely on supersymmetry to cancel the
quadratic divergent corrections have been proposed and studied. The little
Higgs model \cite{Arkani-Hamed:2001nc} is one of these alternatives. In this
model, the Higgs boson is regarded as a pseudo Nambu-Goldstone boson, which
originates from the spontaneous breaking of a global symmetry at certain
high scale, and the global symmetry protects the Higgs mass from the quadratic
radiative corrections. The simplest version of the model is called the littlest
Higgs model \cite{Arkani-Hamed:2002qy}. The global symmetry of the model is
SU(5), which is spontaneously broken into SO(5). The part of the SU(5) symmetry
is gauged, and the gauge symmetry is SU(2)$_1$ $\times$ SU(2)$_2$ $\times$
U(1)$_1$ $\times$ U(1)$_2$. The top sector is also extended to respect the part
of the global symmetry.

This model, however, predicts a large correction to the EW observables because
of the direct mixing between heavy and light gauge bosons after the EW symmetry
breaking. The precision EW measurements force the masses of heavy gauge 
bosons and
top partners to be ${\cal O}$ (10 TeV), reintroducing the fine tuning problem
to the Higgs mass \cite{difficulty}. A solution of the problem is the
introduction of T-parity to the model which forbids the mixing
\cite{Tparity}. This is the symmetry under the transformations, SU(2)$_1$
$\leftrightarrow$ SU(2)$_2$ and U(1)$_1$ $\leftrightarrow$ U(1)$_2$. All heavy
gauge bosons are assigned a T-odd charge, while the SM particles have
a T-even charge. The matter sector should be extended so that T-odd 
partners are
predicted. The lightest T-odd particle is the heavy photon, which is stable
and becomes a candidate for dark matter
\cite{Hubisz:2004ft}-\cite{Birkedal:2006fz}.

Starting from a different underlying theory, the littlest Higgs model with
T-parity ends up predicting a similar phenomenology to that of the MSSM. We
view that this is indispensable for the model to remove the hierarchy problem.
One needs a new symmetry to protect the Higgs mass to reduce the quadratic
divergence of the theory in a meaningful manner. The symmetry must involve both
the top quark and the gauge sectors, because the Higgs couplings to 
these particles are
the dominant source of the divergences. Unless some parity is not assigned to
the gauge partner, large corrections to the EW observables are expected, and
those are not acceptable after the LEP era. Some of the top and gauge partners
are required in the parity odd sector of the model, while SM particles are in
the parity even sector. Notably if this parity is exact, the lightest parity
odd particle is stable, therefore the picture is consistent with the existence
of the dark matter in our Universe. Finally, the production of the top partners
and their decays into the top quark and an invisible particle are 
predicted as the
signal of ``Beyond the Standard Model'' at the LHC.

For the case of the MSSM, the pair production of the scalar top is not
detectable, because the production cross section is too small. 
Supersymmetry is instead detected by the production of gluinos and squarks in
the first generation, and the scalar top may appear as the decay product of the
gluino \cite{Hisano:2003qu}. On the other hand, the cross section of the top
partner pair production in the littlest Higgs model with T-parity is about 10
times larger than that of the scalar top, so that they may be detectable at the
LHC \cite{Meade:2006dw}. The purpose of this paper is to provide a realistic
estimate for the detection of the fermionic top partner at the LHC.

This paper is organized as follows. In Sec.\ref{LstHwT}, we review the littlest
Higgs model with T-parity focusing on the top sector. In Sec.\ref{Constraints},
we summarize the electroweak and dark matter constraints on the model. We find
that the lower limit of the top partner mass is about 600~GeV, and 
the lightest T-odd 
particle is always much lighter than the top partner. The signal at
the LHC is therefore two top quarks with significant transverse momentum and missing
energy.

In Sec.\ref{Signature} and \ref{Discovery}, we discuss the top partner
signature at the LHC. Reconstructing a top quark in the event is essential to
identify the top partner. We describe the method to reconstruct the top quark
in Sec.\ref{Signature}. We apply a hemisphere analysis to the signal
reconstruction, and study the reconstruction efficiency of the top quark for
both signal and $t\bar{t}$ background. In Sec.\ref{Discovery}, we discuss the
basic cuts to reduce the top quark background. The highest S/N ratio is
obtained in the region where the effective mass is around twice the top
partner mass. We also show the numerical results of our simulation study and
find that the top partner signature would be significant over the
background if the mass of the top partner is less than 1 TeV. Sec.\ref{Summary}
is devoted to discussion and comments.

\section{The Littlest Higgs Model with T-parity}
\label{LstHwT}

In this section, we briefly review the littlest Higgs model with T-parity
focusing on the top sector of the model. The constraints on the model from
WMAP observations and electroweak precision measurements will be discussed
in the next section. For general reviews of the little Higgs models and their
phenomenological aspects, see Refs.\cite{littlest_review,littlest}.

\subsection{Gauge-Higgs sector}

The littlest Higgs model \cite{Arkani-Hamed:2002qy} is based on a non-linear
sigma model describing an SU(5)/SO(5) symmetry breaking. The non-linear sigma
field, $\Sigma$, is given as
\begin{eqnarray}
 \Sigma = e^{2i\Pi/f}\Sigma_0~,
\end{eqnarray}
where $f$ is the vacuum expectation value associated with the breaking
and is expected to be ${\cal O}(1)$ TeV. The
Nambu-Goldstone (NG) boson matrix $\Pi$ and the direction of the breaking
$\Sigma_0$ are
\begin{eqnarray}
 \Pi
 =
 \left(
  \begin{array}{ccc}
                   0 & H  /\sqrt{2} & \Phi         \\
   H^\dagger/\sqrt{2} &            0 & H^T/\sqrt{2} \\
         \Phi^\dagger & H^*/\sqrt{2} & 0            \\
  \end{array}
 \right)~,
 \qquad
 \Sigma_0
 =
 \left(
  \begin{array}{ccc}
         0 & 0 & {\bf 1} \\
         0 & 1 &       0 \\
   {\bf 1} & 0 &       0 \\
  \end{array}
 \right)~.
 \label{pNG matrix}
\end{eqnarray}
where we omit would-be NG fields in the $\Pi$ matrix. An
[SU(2)$\times$U(1)]$^2$ subgroup in the SU(5) global symmetry is gauged, which
is broken down to the diagonal subgroup identified with the SM gauge group,
SU(2)$_L\times$U(1)$_Y$. Due to the presence of the gauge interactions (and
Yukawa interactions introduced in the next subsection), the SU(5) global
symmetry is not exact, and particles in the $\Pi$ matrix become pseudo NG
bosons. Fourteen (= 24 $-$ 10) NG bosons are decomposed into representations
under the electroweak gauge group as ${\bf 1}_0 \oplus {\bf 3}_0 \oplus
{\bf 2}_{\pm 1/2} \oplus {\bf 3}_{\pm 1}$. The first two representations are
real, and become longitudinal components of heavy gauge bosons when the
[SU(2)$\times$U(1)]$^2$ is broken down to the SM gauge group. The other scalars
in the representations ${\bf 2}_{\pm 1/2}$ and ${\bf 3}_{\pm 1}$ are the complex
doublet identified with the SM Higgs field ($H$ in Eq.(\ref{pNG matrix})) and a
complex triplet Higgs field ($\Phi$ in Eq.(\ref{pNG matrix})), respectively.
The kinetic term of the $\Sigma$ field is given as
\begin{eqnarray}
 {\cal L}_{\Sigma}
 =
 \frac{f^2}{8}{\rm Tr}
 \left[
  D_\mu\Sigma\left(D^\mu\Sigma\right)^\dagger
 \right]~,
 \label{Kinetic L}
\end{eqnarray}
where
\begin{eqnarray}
 D_\mu\Sigma
 =
 \partial_\mu\Sigma
 -
 i
 \sum_{j = 1}^2
 \left[
  g_j  ({\bf W}_j\Sigma + \Sigma {\bf W}_j^T)
  +
  g'_j ({\bf B}_j\Sigma + \Sigma {\bf B}_j^T)
 \right]~.
\end{eqnarray}
Here, ${\bf W}_j = W^a_j Q_j^a$ (${\bf B}_j = B_j Y_j$) is the SU(2)$_j$
(U(1)$_j$) gauge field and $g_j (g'_j)$ is the corresponding gauge coupling
constant. The generators $Q_j$ and $Y_j$, are
\begin{eqnarray}
 &&
 Q_1^a
 =
 ~~~
 \frac{1}{2}
 \left(
  \begin{array}{ccc}
   \sigma^a & 0 & 0
   \\
   0 & 0 & 0
   \\
   0 & 0 & 0
  \end{array}
 \right)~,
 \qquad
 Y_1
 =
 {\rm diag}(3,3,-2,-2,-2)/10~,
 \nonumber \\
 &&
 Q_2^a
 =
 -\frac{1}{2}
 \left(
  \begin{array}{ccc}
   0 & 0 & 0
   \\
   0 & 0 & 0
   \\
   0 & 0 & \sigma^{a*}
  \end{array}
 \right)~,
 \qquad
 Y_2
 =
 {\rm diag}(2,2,2,-3,-3)/10~,
\end{eqnarray}
where $\sigma^a$ is the Pauli matrix.

In terms of the above fields, the symmetry under T-parity \cite{Tparity} is
defined as the invariance of the Lagrangian under the transformation:
\begin{eqnarray}
 W^a_1 \leftrightarrow W^a_2~,
 \qquad
 B_1 \leftrightarrow B_2~,
 \qquad
 \Pi \leftrightarrow -\Omega\Pi\Omega~
 ({\rm or~}\Sigma \leftrightarrow
  \tilde{\Sigma}
  \equiv
  \Sigma_0\Omega\Sigma^\dagger\Omega\Sigma_0)~,
\end{eqnarray}
where $\Omega = {\rm diag}(1,1,-1,1,1)$. As a result of the symmetry, the
gauge coupling $g_1$ ($g'_1$) must be equal to $g_2$ ($g'_2$), namely $g_1 =
g_2 = \sqrt{2}g$ ($g'_1 = g'_2 = \sqrt{2}g'$), where $g$ ($g'$) is nothing but
the coupling constant of the SM SU(2)$_L$ (U(1)$_Y$) gauge symmetry.

The Higgs potential is generated radiatively
\cite{Arkani-Hamed:2002qy,Hubisz:2004ft}
\begin{eqnarray}
 V(H, \Phi)
 =
 \lambda f^2{\rm Tr}\left[\Phi^\dagger\Phi\right]
 -
 \mu^2H^\dagger H
 +
 \frac{\lambda}{4}\left(H^\dagger H\right)^2
 +
 \cdots~.
 \label{Higgs Potential}
\end{eqnarray}
The main contributions to $\mu^2$ come from the logarithmic divergent corrections
at 1-loop level and quadratic divergent corrections at 2-loop level. As a
result, $\mu^2$ is expected to be smaller than $f^2$. The triplet Higgs mass
term, on the other hand, receives quadratic divergent corrections at 1-loop level,
and therefore is proportional to $f^2$. The quartic coupling $\lambda$ is determined
by the 1-loop effective potential from the gauge and top sectors. Since both $\mu$
and $\lambda$ depend on the parameters at the cutoff scale, we treat them as free
parameters in this paper.

Next, we discuss the mass spectrum of the gauge and Higgs bosons. This model
contains four kinds of gauge fields, $W^a_1$, $W^a_2$, $B_1$ and $B_2$, in the
electroweak gauge sector. The linear combinations $W^a = (W^a_1 + W^a_2)/\sqrt{2}$
and $B = (B_1 + B_2)/\sqrt{2}$ correspond to the SM gauge bosons for the
SU(2)$_L$ and U(1)$_Y$ symmetries. The other linear combinations, $W^a_H = (W^a_1 -
W^a_2)/\sqrt{2}$ and $B_H = (B_1 - B_2)/\sqrt{2}$, are additional gauge bosons,
which acquire masses of ${\cal O}(f)$ through the SU(5)/SO(5) symmetry
breaking. After the electroweak symmetry breaking, the neutral components of $W^a_H$
and $B_H$ are mixed with each other, and form mass eigenstates $A_H$ and $Z_H$. The
masses of the heavy bosons are
\begin{eqnarray}
 m_{A_H}
 \simeq
 0.45 g' f~,
 \qquad
 m_{Z_H}
 \simeq
 m_{W_H}
 \simeq
 g f~.
 \label{AH mass}
\end{eqnarray}
The mixing angle $\theta_H$ between $W^3_H$ and $B_H$ is considerably 
suppressed.  It is given by
$\tan\theta_H \simeq - g' v^2/(4g f^2)$, where $v$ ($\simeq$ 246~GeV) is the
vacuum expectation value of the Higgs field. Thus the dominant component 
of $A_H$ is $B_H$. Finally, the
mass of the triplet Higgs boson $\Phi$ is given by $m_\Phi^2 = \lambda f^2 =
2m_h^2f^2/v^2$, where $m_h$ is the SM Higgs boson mass.

Under T-parity, the new heavy gauge bosons and the triplet Higgs boson behave
as T-odd particles, while SM particles are T-even. As shown in
Eq.(\ref{AH mass}), the heavy photon is considerably lighter than
the other T-odd particles. Stability of $A_H$ is guaranteed by T-parity conservation,
and it becomes a candidate for dark matter.

\subsection{Top sector}

To implement T-parity, two SU(2) doublets, $q_1$ and $q_2$, are introduced
for each SM fermion doublet. Furthermore, two vector-like singlet top partners,
$U_1$ and $U_2$, are also introduced in the top sector in order to cancel large
radiative corrections to the Higgs mass term. Since we are interested in top
partner production at the LHC, only the top sector is discussed here. For the
other matter sectors, see Refs.\cite{Hubisz:2004ft,Belyaev:2006jh}.

\begin{table}[t]
\center{
 \begin{tabular}{|c|c||c|c|}
  \hline
    $q_1$ & $({\bf 2}, 1/30; {\bf 1}, 2/15)$ &
    $q_2$ & $({\bf 1}, 2/15; {\bf 2}, 1/30)$ \\
  \hline
  $U_{L1}$ & $({\bf 1}, 8/15; {\bf 1}, 2/15)$ &
  $U_{L2}$ & $({\bf 1}, 2/15; {\bf 1}, 8/15)$ \\
  \hline
  $U_{R1}$ & $({\bf 1}, 8/15; {\bf 1}, 2/15)$ &
  $U_{R2}$ & $({\bf 1}, 2/15; {\bf 1}, 8/15)$ \\
  \hline
    $u_R$ & $({\bf 1}, 1/3 ; {\bf 1}, 1/3 )$ &
          &                                  \\
  \hline
 \end{tabular}
 }
\caption{\small The [SU(2)$\times$ U(1)]$^2$ charges for particles in the top
         sector.}
\label{charges}
\end{table}

The quantum numbers of the particles in the top sector under 
the [SU(2)$\times$ U(1)]$^2$ gauge symmetry are shown 
in Table \ref{charges}.  All fields in the table are 
triplets under the SM SU(3)$_c$ (color) symmetry. Using these
fields, the Yukawa interaction terms which are invariant under T-parity and
gauge symmetries turn out to be
\begin{eqnarray}
 {\cal L}_t
 &=&
 -\frac{\lambda_1 f}{2\sqrt{2}}
 \epsilon_{ijk}\epsilon_{xy}
 \left[
  (\bar{{\cal Q}}_1)_i \Sigma_{j x} \Sigma_{k y}
  -
  (\bar{{\cal Q}}_2 \Sigma_0)_i \tilde{\Sigma}_{j x} \tilde{\Sigma}_{k y}
 \right]u_R
 \nonumber \\
 &&
 -
 \lambda_2 f
 \left(
  \bar{U}_{L1}U_{R1}
  +
  \bar{U}_{L2} U_{R2}
 \right)
 +
 h.c.~,
\end{eqnarray}
where
\begin{eqnarray}
 &&
 {\cal Q}_i
 =
 \left(
  \begin{array}{c}
   q_i \\ U_{Li} \\ 0
  \end{array}
 \right)~,
 \qquad
 q_i
 =
 -\sigma^2
 \left(
  \begin{array}{c}
   u_{Li} \\ b_{Li}
  \end{array}
 \right)~.
\end{eqnarray}
The indices $i,j,k$ run from 1 to 3 whereas $x,y = 4,5$. The coupling constant
$\lambda_1$ is introduced as the top Yukawa coupling, while $\lambda_2 f$ gives
the vector-like mass term for the singlet fields. Under T-parity, $q_i$ and
$U_i$ transform as $q_1 \leftrightarrow - q_2$ and $U_1 \leftrightarrow - U_2$.
Therefore, the T-parity eigenstates are given by
\begin{eqnarray}
 q_\pm
 =
 \frac{1}{\sqrt{2}}(q_1 \mp q_2)~,
 \qquad
 U_{L\pm}
 =
 \frac{1}{\sqrt{2}}(U_{L1} \mp U_{L2})~,
 \qquad
 U_{R\pm}
 =
 \frac{1}{\sqrt{2}}(U_{R1} \mp U_{R2})~.
\end{eqnarray}

In terms of these eigenstates, the mass terms for these quarks are written as
follows,
\begin{eqnarray}
 {\cal L}_{\rm mass}
 =
 -
 \lambda_1
 \left[
  f\bar{U}_{L+}
  +
  v\bar{u}_{L+}
 \right]u_R
 -
 \lambda_2 f
 \left(
  \bar{U}_{L+} U_{R+}
  +
  \bar{U}_{L-} U_{R-}
 \right)
 +
 h.c.~.
\end{eqnarray}
The remaining T-odd fermion, $q_-$, acquires mass by introducing
an additional SO(5) multiplet transforming nonlinearly under the SU(5) symmetry.
Therefore, the mass term of the $q_-$ quark does not depend on $\lambda_1$ and
$\lambda_2$. In this paper, we assume that the $q_-$ quark is heavy 
compared to other top partners, and do not consider its production at the LHC.
For the $q_-$ quark phenomenology, see Ref.\cite{Belyaev:2006jh}.

The T-even states $u_+$ and $U_+$ form the following mass eigenstates
\begin{eqnarray}
 &&
 t_L
 =
 \cos\beta~u_{L+} - \sin\beta~U_{L+}~,
 \qquad
 T_L
 =
 \sin\beta~u_{L+} + \cos\beta~U_{L+}~,
 \nonumber \\
 &&
 t_R
 =
 \cos\alpha~u_R - \sin\alpha~U_{R+}~,
 \qquad
 T_R
 =
 \sin\alpha~u_R + \cos\alpha~U_{R+}~,
\end{eqnarray}
where $\sin\alpha\simeq\lambda_1/(\lambda_1^2 + \lambda_2^2)^{1/2}$, and
$\sin\beta\simeq\lambda_1^2v/[(\lambda_1^2 + \lambda_2^2)f]$. The $t$ quark is
identified with the SM top quark, and $T$ is its T-even heavy partner. On the
other hand, the T-odd fermions $U_{L-}$ and $U_{R-}$ form a Dirac 
fermion, $T_-$.
The masses of these quarks are given by
\begin{eqnarray}
 m_t
 =
 \frac{\lambda_1\lambda_2v}{\sqrt{\lambda_1^2 + \lambda_2^2}}~
 \qquad
 m_T
 =
 \sqrt{\lambda_1^2 + \lambda_2^2}f~,
 \qquad
 m_{T_-}
 =
 \lambda_2f~.
 \label{t_mass}
\end{eqnarray}
It is worth noting that the T-odd states do not participate in the
cancellation of quadratic divergent corrections to the Higgs mass term. The
cancellation is achieved only by loop diagrams involving $t$ and $T$ quarks.

\section{WMAP and EW precision constraints}
\label{Constraints}

In the previous section, four parameters are introduced in the
gauge-Higgs and top
sectors in addition to the gauge coupling constants ($g$ and $g'$) and 
the vacuum
expectation value of the Higgs field ($v$). Those are $m_h$, $f$, $\lambda_1$,
and $\lambda_2$. Since the top quark mass is determined by the combination of
$v$, $\lambda_1$ and $\lambda_2$, the number of undetermined parameters is
three. These parameters can be expressed by $m_h$, $m_{A_H}$, and $m_{T_-}$.

In this section, we see that $m_h$ and $m_{A_H}$ are directly related each other
thanks to the precise data of the WMAP observations. Therefore, it is possible
to write down the Higgs mass as a function of the dark matter mass, $m_h =
m_h(m_{A_H})$. The electroweak precision measurements provide a further
constraint on $m_h$ and $m_{T_-}$. They gives a lower bound on $m_{T_-}$ and
a large mass difference between $T_-$ and $A_H$.

\subsection{Constraint from WMAP observation}

First, we consider the WMAP constraint on the littlest Higgs model with
T-parity. The dark matter, $A_H$, annihilates mainly into weak gauge bosons,
$W^+W^-$ and $ZZ$ through diagrams in which the Higgs boson propagates in the
s-channel. Once we calculate the annihilation cross section, the thermal relic
abundance of the dark matter is obtained by solving the Boltzmann equation.
For the detailed calculation of the abundance in this model, see
Refs.\cite{Hubisz:2004ft}-\cite{Birkedal:2006fz}.  To good accuracy, the
dark matter abundance can be written as
\begin{eqnarray}
 \Omega_{\rm DM} h^2
 =
 8.4\times 10^{-2}
 \left(
  \frac{1{\rm pb}\cdot c}{\sigma v_{\rm rel}}
 \right)~,
\end{eqnarray}
where $c$ is the speed of light, and $v_{\rm rel}$ is the relative velocity between initial
dark matters. Since the product of the cross section and the relative
velocity, $\sigma v_{\rm rel}$, and hence the dark matter
abundance, $\Omega_{\rm DM}h^2$, depend only on $m_{A_H}$ and $m_h$, the WMAP
observation, $\Omega_{\rm DM} h^2 \simeq$ 0.111, gives a relation between these
parameters.

\begin{figure}[t]
\begin{center}
\scalebox{0.6}{\includegraphics*{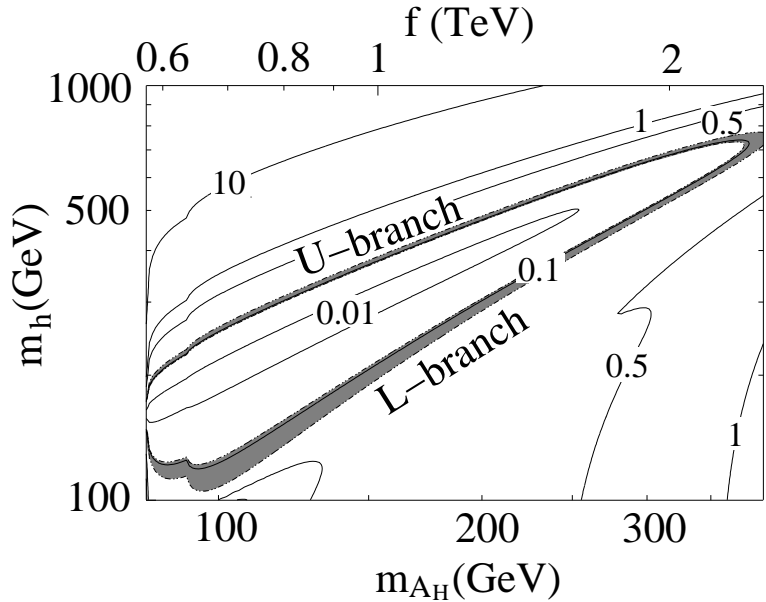}}
\caption{\small Contour plot of the thermal relic abundance of the dark matter,
         $\Omega_{\rm DM} h^2$, in the $(m_{A_H},m_h)$ plane. The thin shaded area is
         the allowed region from the WMAP observation at 2$\sigma$ level,
         $0.094 < \Omega_{\rm DM} h^2 < 0.129$.}
\label{fig:abundance}
\end{center}
\end{figure}

In Fig.\ref{fig:abundance}, the thermal relic abundance of dark matter is
depicted as a contour plot in the $(m_{A_H},m_h)$ plane. The thin shaded area is the
allowed region from the WMAP observation at 2$\sigma$ level, $0.094 < 
\Omega_{\rm DM}
h^2 < 0.129$ \cite{Spergel:2003cb}. In the figure, there are two branches: the
upper branch (U-branch) and the lower branch (L-branch). These branches are
sometimes called ``Low'' and ``High'' regions, respectively
\cite{Birkedal:2006fz}. In the U-branch, the Higgs boson mass is larger than
twice the dark matter mass, $m_h > 2m_{A_H}$, while $m_h < 2m_{A_H}$ in the
L-branch. The Higgs mass is precisely determined by the dark matter mass
up to a
two-fold ambiguity by imposing the WMAP constraint.

\subsection{Constraint from electroweak precision measurement}

Next, we discuss the constraint from electroweak precision measurements.
New physics contributions to electroweak observables come from radiative
corrections, since there is no tree-level effect due to T-parity. The
constraint is sensitive to the masses of the Higgs boson and the top partner. Since the Higgs
mass is directly related to the dark matter mass through the WMAP constraint,
the electroweak precision constraint on $m_h$ and $m_{T_-}$ is translated into
one on $m_{A_H}$ and $m_{T_-}$.

In order to obtain the constraint, we follow the procedure in
Ref.\cite{Hubisz:2005tx} using the S, T and U parameters \cite{Peskin:1991sw}. In
that paper, it has been shown that main contributions to the parameters come
from the top-sector
and the custodial-symmetry violating effect
from heavy gauge boson loops. However, in Ref.\cite{Asano:2006nr}, it has
been shown that the latter contribution is negligibly small compared to the
former one. Therefore, we consider only the top sector contribution in order to
obtain the constraint. For the detailed expression of the top sector
contributions, see Ref.\cite{Hubisz:2005tx}.

\begin{figure}[t]
\begin{center}
\scalebox{0.28}{\includegraphics*{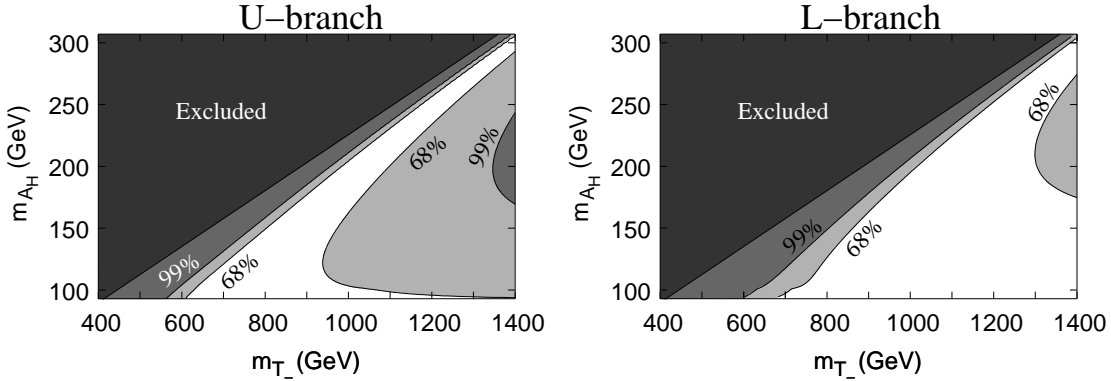}}
\caption{\small Constraints on $m_{T_-}$ and $m_{A_H}$ at 68\% and 99\%
         confidence level. At each point in these figures, the Higgs mass
         is determined to satisfy the WMAP constraint on the U-branch (left
         figure) and L-branch (right figure).}
\label{fig:mT_range}
\end{center}
\end{figure}

In Fig.\ref{fig:mT_range}, constraints on $m_{T_-}$ and $m_{A_H}$ at 68\% and
99\% confidence level are depicted. At each point in these figures, the Higgs
mass is determined to satisfy the WMAP constraint on the U-branch (left figure)
and L-branch (right figure). To obtain the constraint, we have used three
experimental values:
the $W$ boson mass ($m_W = 80.412 \pm 0.042$~GeV) and 
the weak mixing angle ($\sin^2\theta^{\rm lept}_{\rm eff} = 0.23153 \pm 0.00016$) \cite{EWPO method}, 
and the leptonic width of the $Z$ boson ($\Gamma_l =
83.985\pm 0.086$~MeV) \cite{:2005em}. We have also used the fine structure
constant at the $Z$ pole ($\alpha^{-1}(m_Z) = 128.950\pm 0.048$) and the top
quark mass ($m_t = 172.7\pm 2.9$~GeV) \cite{unknown:2005cc}.

As seen in these figures, the lower bound of the $T_-$ quark mass is about
600~GeV, and the mass difference between $m_{T_-}$ and $m_{A_H}$ is larger than
500~GeV\footnote{As shown in the previous subsection, the dominant annihilation
mode of dark matter for the relic abundance is $A_HA_H \rightarrow W^+W^-$,
implying that $m_{A_H} > m_W$.}. In the excluded region painted black,
there is no combination of the parameters $\lambda_1, \lambda_2$ and $f$
which can give the correct top quark mass via Eq.~(\ref{t_mass}).
The parameter
region with large $m_{A_H}$ corresponds to the region with heavy $m_h$ due to the
WMAP constraint. The heavy Higgs contribution to the T-parameter can be cancelled by
those from the top partner, $T$, if its mass is tuned appropriately. When
$m_{A_H} \sim m_h/2$ increases, the cancellation can be achieved only for small
region of $m_{T_-}$ as we can see in these figures.

\section{$T_-$ quark production at the LHC} 
\label{Signature}

In this section, we study the signature of $T_-$ quark production at the
LHC. Following the discussion in the previous section, we first discuss
properties of the $T_-$ quark and present a few representative points used in
our simulation study. We next consider a top reconstruction for the signal
using a hemisphere analysis. Finally, we address the background to
this process, which comes from SM $t\bar{t}$ production.

\subsection{Properties of the $T_-$ quark}

\begin{figure}[t]
\begin{center}
\scalebox{0.42}{\includegraphics*{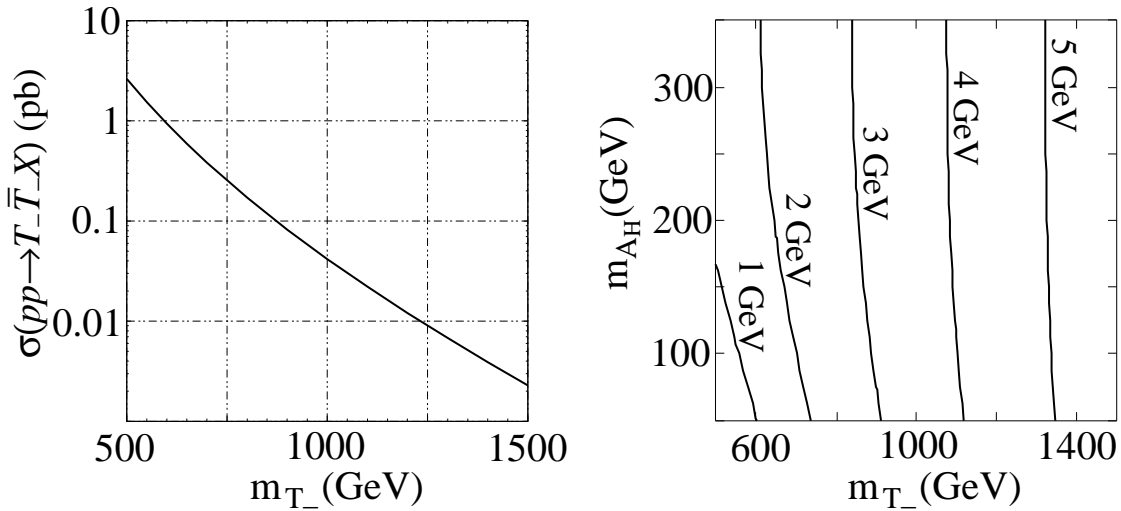}}
\caption{\small Production cross section of the $T_-$ quark at the LHC as a
         function of $m_{T_-}$ (left figure). Contour plot of the $T_-$ decay
         width in the ($m_{T_-},m_{A_H}$) plane (right figure).}
\label{fig:prop_Tm}
\end{center}
\end{figure}

Due to the T-parity conservation, the $T_-$ quark would be pair produced from
proton-proton ($pp$) collisions at the LHC. Since the $T_-$ quark has a color
charge, it is produced dominantly through SU(3)$_c$ interactions at the LHC. The
production cross section depends only on its mass, $m_{T_-}$. Unlike the scalar top
in the MSSM, the $T_-$ quark is a Dirac fermion. Hence its production cross
section at the LHC ranges between 0.1-1 pb when $m_{T_-}$ is less than 1 TeV as
shown in Fig.\ref{fig:prop_Tm} (left figure).

The decay process of the $T_-$ quark is simple, because only $A_H$ and $Z_H$
($W_H$) are T-odd particles lighter than the $T_-$ quark. Furthermore, the
$T_-$ quark is almost an SU(2)$_L$ singlet. The interactions relevant to the decay are
\begin{eqnarray}
 {\cal L}
 &=&
 ~~~
 i\frac{2g'}{5}\cos\theta_H \bar{T}_-\Slash{A}_H
 \left(\sin\beta P_L + \sin\alpha P_R\right)t
 \nonumber \\
 &&
 +~
 i\frac{2g'}{5}\sin\theta_H \bar{T}_-\Slash{Z}_H
 \left(\sin\beta P_L + \sin\alpha P_R\right)t~.
 \label{decay_int}
\end{eqnarray}
As seen in equation (\ref{decay_int}), the decay mode $T_- \rightarrow
Z_Ht$ is highly suppressed by $\sin^2\theta_H$, therefore $T_-$ decays
dominantly into the stable $A_H$ and the top quark. In Fig.\ref{fig:prop_Tm}
(right figure), the decay width is depicted as a contour plot in the ($m_{T_-},
m_{A_H}$) plane. It shows that the width is typically several GeV.

The signal of $T_-$ quark production is a top pair ($t\bar{t}$) with
significant missing transverse momentum. In this paper, we assume that the
branching ratio of the process $T_- \rightarrow A_H t$ is 100\%. For
simplicity we also assume that the other extra matter fermions do not contribute 
to the signal. The model points we choose for the simulation study are listed in Table
\ref{point}. The production cross sections are also shown in the table, which were
obtained by the CompHep code \cite{Boos:2004kh}  using the ``CTEQ6L1'' parton
distribution function \cite{Kretzer:2003it} and the scale of the QCD
coupling set to be $Q=m_{T_{-}}$\footnote{The cross sections are a factor of
2 smaller
than those listed in \cite{Meade:2006dw}. We found their cross section is
obtained at the scale of $m_Z$.}.

\begin{table}[t]
\center{
\begin{tabular}{|c||c|c|c|}
 \hline
  & $m_{T_-}$ (GeV) & $m_{A_H}$ (GeV) &
  $\sigma(pp\rightarrow T_-\bar{T}_- + X)$ (pb)\\
 \hline
 \hline
    I & 600 & 100 & 0.940 \\
 \hline
   II & 700 & 125 & 0.382 \\
 \hline
  III & 800 & 150 & 0.171 \\
 \hline
   IV & 900 & 175 & 0.0822 \\
 \hline
\end{tabular}
 }
\caption{\small The model points for our simulation study. The production cross
         section for the $T_-$ quark is also shown.}
\label{point}
\end{table}

In order to generate parton level events, we calculated the process $pp
\rightarrow t \bar{t} A_H A_H$ directly using the CompHep code, keeping
diagrams relevant to the process through on-shell $T_-\bar{T}_-$ production.
We generated 100,000 events for each model point to study their distributions.
The parton level events were interfaced to HERWIG \cite{Corcella:2000bw} for
fragmentation, initial and final state radiations, and hadronization. The
effect of the top polarization is not included in our simulation. The detector
effects were simulated by the AcerDET code \cite{Richter-Was:2002ch}. This code
provides a simple detector simulation and jet reconstruction using a simple cone
algorithm. It also identifies isolated leptons and photons, finds $b$ and
$\tau$ jets, and calculates the missing momentum of the events using calorimeter
information.

Before going to the discussion of top reconstruction, we define two important
quantities which are frequently used for new physics searches at the LHC. 
One is the missing transverse energy, $E_{\rm Tmiss} \equiv (P_{T x}^2 +
P_{T y}^2)^{1/2}$, which is important for signal of the models with a stable dark
matter candidate such as the MSSM or the little Higgs model with T-parity. Here
$P_{T i}$ is the sum of transverse momenta measured by the calorimeter. The
other is the effective transverse mass,
\begin{equation}
 M_{\rm eff}
 \equiv
 \sum_{\rm jets} p_T
 +
 \sum_{\rm isolated~leptons} p_T
 +
 \sum_{\rm isolated~photons} p_T
 +
 E_{\rm Tmiss}~,
\end{equation}
where we require that the pseudo rapidity is less than three ($\eta<3$) and the
transverse momentum is $p_T > 50~(10)$~GeV for each jet (lepton/photon). The effective
transverse mass is a good quantity to measure the mass scale of a produced
particle.

\subsection{Top reconstruction} 

Now we move on to the discussion of top quark reconstruction for the
signal. The top
quarks produced from $T_-$ quarks have significant $p_T$. Since the $T_-$ quarks
must be produced in a pair, we expect two separate jet 
systems  
originating from the 
two top quarks. We therefore use the hemisphere analysis \cite{Filip} for the
event reconstruction. Two hemispheres are defined in each event, and high $p_T$
jets, leptons and photons are assigned to one of the hemispheres. 
Specifically, 
\begin{itemize}
\item
  Each hemisphere is defined by an axis $P_i$ $(i = 1, 2)$, which is the sum of
  the momenta of high $p_T$ objects (jets/leptons/photons) belonging to the
  hemisphere $i$. Only the jets with $p_T > 50$~GeV and leptons/photons with
  $p_T > 10$~GeV are assigned to the hemisphere in order to reduce the
  contamination of QCD activity.
\item
  High $p_T$ objects $k$ belonging to the hemisphere $i$ satisfy the following
  condition
  \begin{equation}
    d(p_k, P_i) < d(p_k,P_j)~,
  \end{equation}
  where $i$ and $j$ are the indices of the hemispheres, and the function $d$ is
  defined as
  \begin{equation}
    d(p_k, P_i)
    =
    \left(E_i - \left|P_i\right|\cos\theta_{i k}\right)
    \frac{E_i}{(E_i + E_k)^2}~,
  \end{equation}
  where $\theta_{i k}$ is the angle between $P_i$ and $p_k$.
\end{itemize}

To find the axis satisfying the above conditions, we take following steps. (1)
We take the highest $p_T$ object $i$ (jets/leptons/photons) with momentum
$p_i$, and the object $j$ with largest $\Delta R |p_j|$, where $\Delta R =
[(\Delta\phi(i,j))^2 + (\Delta\eta(i,j))^2]^{1/2} $. We take $p_i$ and $p_j$ to
be the seeds of the hemisphere axes, namely, $P^{\rm in}_1 = p_i$, $P^{\rm in}_2 =
p_j$. (2) Each object with momentum $p_k$ is assigned to the hemisphere $i$, if
$d(p_k, P^{\rm in}_i) < d(p_k, P^{\rm in}_j)$. (3) We then define new
$P^{\rm in}_i$ ($i = 1, 2$) as the sum of the momenta of the objects belonging
to the hemisphere $i$. (4) We repeat the processes (2) and (3) until assignment
converges. In this paper, we denote the hemisphere seeded from the highest
$p_T$ object as ``hemisphere 1'' and the other as ``hemisphere 2''.

\begin{figure}[t]
\begin{center}
\scalebox{0.35}{\includegraphics*{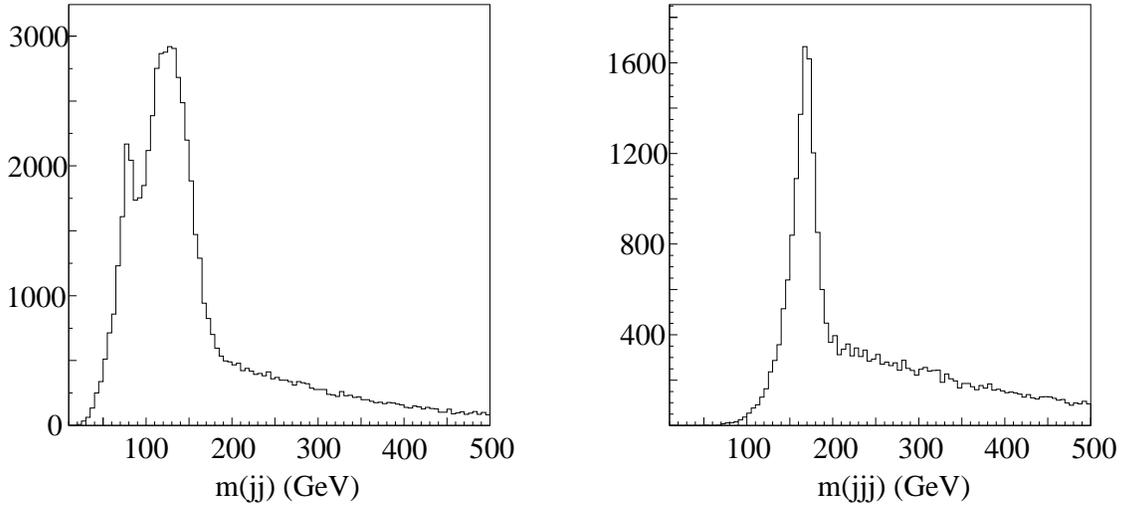}}
\caption{\small Distribution of a two jet invariant mass, $m_{\rm max}(j j)$
         (left figure), and that of $ m_{\rm min}(j j j)$ (right figure) in the
         $pp \rightarrow T_-\bar{T}_- \rightarrow t\bar{t} A_H A_H$ process
         for point III}
\label{Fig:rjjj}
\end{center}
\end{figure}

If there are more than two jets in a hemisphere, we can calculate the maximum
of the two jet invariant masses for all combinations of jets in the
hemisphere, $m_{\rm max} (j j)$. In Fig.\ref{Fig:rjjj} (left figure), we plot
this for the hemisphere 1 using the point III in Table \ref{point}. The
distribution has two peaks around $m(j j) \sim 80$~GeV and $130$~GeV. The lower
peak corresponds to the jet pair arising from the $W$ decay, while the second
peak comes from the combination of one of the two jets from a $b$ quark and one of the
partons from the $W$ decay.

When a hemisphere contains more than three jets, we can also define the minimum
three jet invariant mass in the hemisphere, $m_{\rm min}(j j j)$, 
where two of the three jets are those which give $m_{\rm max}(j j)$. 
In Fig.\ref{Fig:rjjj} (right figure), the distribution peaks at the input top
quark mass 175~GeV, clearly showing that the hemisphere analysis can group the
jets from the top quark correctly. In the following, we often require the 
``top mass'' cut, $m_{\rm min}(j j j) < 200$~GeV for at least one of the
hemispheres.

The efficiency to find at least one  top quark candidate in a signal event is moderate,
about $20\%$. In Fig.\ref{Fig:rjjj}, we have a long tail at
$m_{\rm min}(j j j) > 200$~GeV, which consists of events with additional
jets or miss-assignment of jets. If we optimize the top search strategy after
the hemisphere reconstruction, we may increase the top reconstruction efficiency,
but we do not study this possibility in this paper. It should be noted that the
reconstructed $m_{\rm min}(j j j)$ distribution in the hemisphere analysis is
not biased, because we do not assume the existence of the top quark in the
event.

\subsection{Background}

The most important background comes from the $pp \rightarrow t\bar{t} + X$ process
in the SM. The tree level production cross section of the top
quark is 400~pb\footnote{The NLO cross section is 800~pb \cite{topNLO}.}. We
have generated events corresponding to 50~fb$^{-1}$ for this study. The
cross section after very weak cuts, $M_{\rm eff} > 400$~GeV, $E_{\rm Tmiss} > 100$
~GeV, $n_{\rm jets}(p_T > 100~{\rm GeV}) \equiv n_{100} \geq 1$ and
$n_{\rm jets}(p_T > 50~{\rm GeV}) \equiv n_{50} \geq 2$ is about 16~pb, which is
still higher than the signal cross section by more than a factor of 10.

\begin{table}[t]
\center{
\begin{tabular}{|c||c|c|}
 \hline
 & $pp \rightarrow T_-\bar{T}_-$ & $pp \rightarrow t\bar{t}$ \\
 \hline
 \hline
 $m_{\rm min}(j j j)_1 < 200$~GeV  or $m_{\rm min}(j j j)_2 < 200$~GeV 
 & 22.9\% & 15.3\% \\
 \hline
 $m_{\rm min}(j j j)_1 < 200$~GeV and $m_{\rm min}(j j j)_2 < 200$~GeV
 &  1.6\% & 0.17\% \\
 \hline
\end{tabular}
 }
\caption{\small Fraction of $T_{-}\bar{T}_{-}$ and $t\bar{t}$ events with
         $m_{\rm min}(j j j) < 200$~GeV in one of the hemispheres, and both of the
         hemispheres. The probability to reconstruct the top quark in both of
         the hemisphere is significantly small for the $t\bar{t}$ production.}
\label{Table:tt_eff}
\end{table}
 
\begin{figure}[t]
\begin{center}
\scalebox{0.32}{\includegraphics*{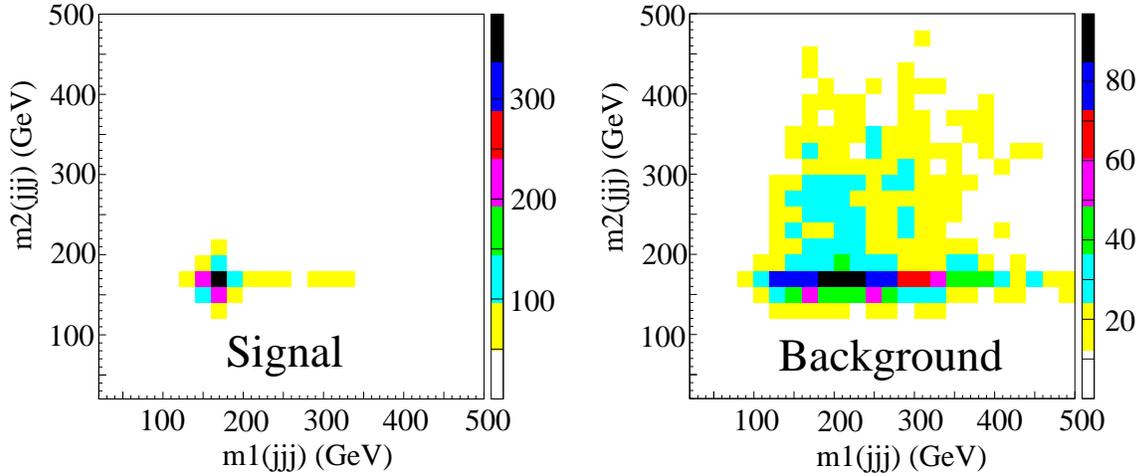}}
\caption{\small Reconstructed $m_{\rm min}(j j j)$ distributions for $pp
         \rightarrow T_-\bar{T}_-$ for point III (left figure) and $pp
         \rightarrow t\bar{t}$ (right figure) processes. We take $E_{\rm Tmiss} >
         400$~GeV and $n_{100} \geq 1$ and $n_{50} \geq 2$ for the $t\bar{t}$
         events. The efficiency to find the top candidate in each hemisphere
         ($m_{\rm min}(j j j)_i < 200$~GeV) is low in the $t\bar{t}$ production compared
         to the signal process.}
\label{mjjj}
\end{center}
\end{figure}

The efficiency to find the top quark candidate in a hemisphere becomes lower for
the $t\bar{t}$ production process if large $E_{\rm Tmiss}$ is required. This is
because the $W$ boson from the top quark decay must decay leptonically to
give such a high $E_{\rm Tmiss}$ to the event. This can be seen in Table
\ref{Table:tt_eff}, where we have listed the probability to find three jets
with $m_{\rm min }(j j j) < 200$~GeV in one, or both, of the hemispheres.
In Fig.\ref{mjjj}, we present the $m_{\rm min}(j j j)$ distribution in
a two dimensional plot, where the x (y) axis corresponds to $m_{\rm min}(j j j)$ in
hemisphere 1 (hemisphere 2). While the distribution peaks around $m_t$ in
both of the hemispheres for the signal (left figure), the distribution for
$t\bar{t}$ production scatters over the plot (right figure). This means that
at least one of the top quarks has to decay leptonically, and therefore some of
the jets must be additional  QCD jets in $t\bar{t}$ production.

If we require a top candidate in both hemispheres, the $t\bar{t}$ background
reduces to the 30 fb level, however the reconstruction efficiency for the signal
also decreases due to additional jets in the final state and the overlap of jets.
In the next section, we look for the excess of signal events
over the $t\bar{t}$ background for the events with $m_{\rm min}(j j j) < 200$~GeV 
in at least one of the hemispheres, for the various 
different sample points. In this case, the $t\bar{t}$
background cross section is around 16~pb $\times$ 0.153 $\sim$ 2.5~pb, which is
many orders of magnitudes larger than any other irreducible backgrounds listed
in \cite{Meade:2006dw}.

In Ref.\cite{Asai}, it has been shown that the quark production is not the
dominant background in the inclusive study of SUSY processes for the 0-lepton +
$E_{\rm Tmiss}$ channel. The backgrounds from $W$ and $Z$ boson production are about
as important as that due to $t\bar{t}$ production. However, one should be able
to reduce these backgrounds significantly by requiring a reconstructed top and
$b$ tagged jets. We therefore do not study these in this paper.

\section{Discovery of the $T_-$ quark at the LHC}
\label{Discovery}

In this section, we investigate the possibility to find the $T_-$ quark at the
LHC. First, we discuss the separation of the signal from the background using
their different kinematic properties. We find the
region in the $E_{\rm Tmiss}$ and $M_{\rm eff}$ plane where the
signal best dominates the background. Next, we calculate the statistical
significance for $T_-$ quark discovery at the LHC, and show that the
significance exceeds seven for all sample points. We also calculate the $M_{T2}$
variable \cite{Barr:2003rg} to investigate the possibility to extract
information about $m_{T_-}$ from the signal. Finally, we comment on the differences
in production and decay distributions between $T_-$ and scalar top signals.

\subsection{Separation of  signals from backgrounds}

\begin{figure}[t]
\begin{center}
\scalebox{0.3}{\includegraphics*{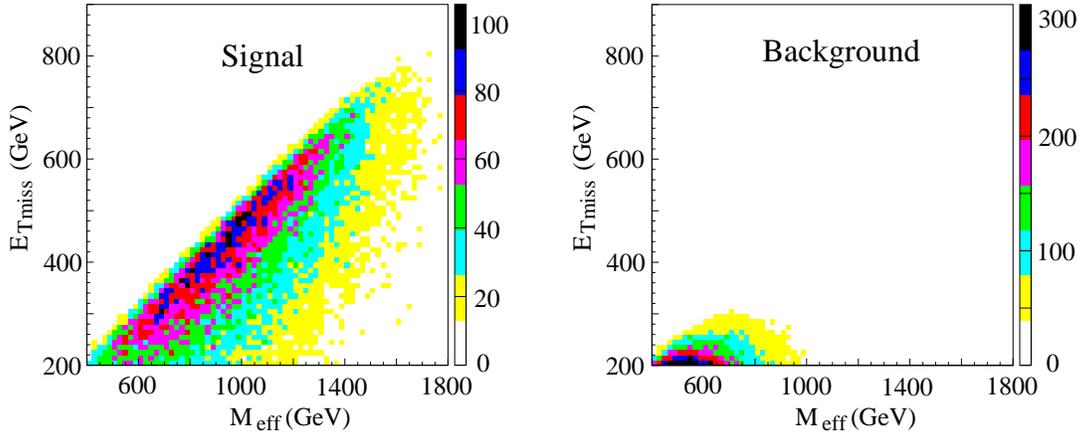}}
\caption{\small $E_{\rm Tmiss}$ versus $M_{\rm eff}$ distributions for the
         $T_-\bar{T}_-$ production at the LHC for point III (left figure), and
         the $t\bar{t}$ background (right figure). Normalizations in both
         figures are arbitrary.}
\label{etmeff}
\end{center}
\end{figure}

\begin{figure}[t]
\begin{center}
\scalebox{0.55}{\includegraphics*{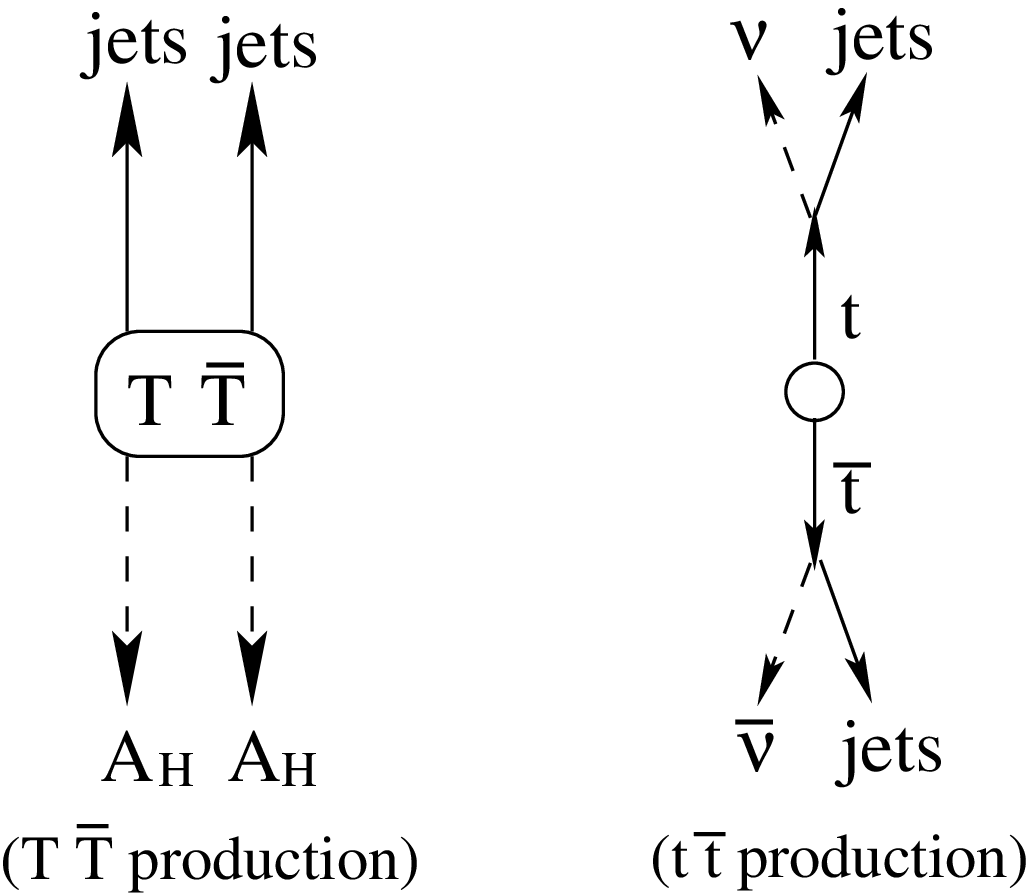}}
\caption{\small $T_-\bar{T}_-$ production which gives the largest $E_{\rm Tmiss}$
         (left figure), and background $t\bar{t}$ production with neutrino
         emissions.}
\label{fig:ETmissMeff}
\end{center}
\end{figure}

In order to separate $T_-\bar{T}_-$ signals from $t\bar{t}$ backgrounds, we
further study $E_{\rm Tmiss}$ distributions for a given (high) $M_{\rm eff}$ interval.
In Fig.\ref{etmeff}, we show the signal and background distributions in the
$M_{\rm eff}$ and $E_{\rm Tmiss}$ plane. The $E_{\rm Tmiss}$ distribution of the signal
peaks near its maximum ($\sim 0.5 M_{\rm eff}$) for significantly large $M_{\rm
eff}$. This feature is common in processes where new particles are pair produced
and each decays into a stable neutral particle and other
visible particles \cite{Kawagoe:2006sm}. This can be understood as follows.
Since new heavy particles are produced mostly near the threshold at the LHC,
the velocity of the $T_-$ quark is low in the transverse direction. Then,
$E_{\rm Tmiss}$ and $M_{\rm eff}$ are maximized when the decay configuration 
of the two $T_-$ quarks is such that the two top quarks from the decay go in the same
direction in the rest frame of the $T_-\bar{T}_-$ system, 
as illustrated in Fig.\ref{fig:ETmissMeff}. In this case, $M_{\rm eff} \sim 2E_{\rm Tmiss}
\sim 2m_{T_-}$.

On the other hand, the kinematics is totally different for the background
$t \bar{t}$ distribution. As shown in Fig.\ref{fig:ETmissMeff}, neutrinos
arising from top decays are collinear to the direction of the parent top quark,
if the center of mass energy in the collision is much higher than the top quark
mass. Therefore, $E_{\rm Tmiss} \sim 0.5 M_{\rm eff}$ is kinematically disfavored.
This can be seen in Fig.\ref{etmeff} (right figure). As $M_{\rm eff}$ increases,
the fraction $E_{\rm Tmiss}/M_{\rm eff}$ is significantly reduced.

\begin{figure}[p]
\begin{center}
\scalebox{0.2}{\includegraphics*{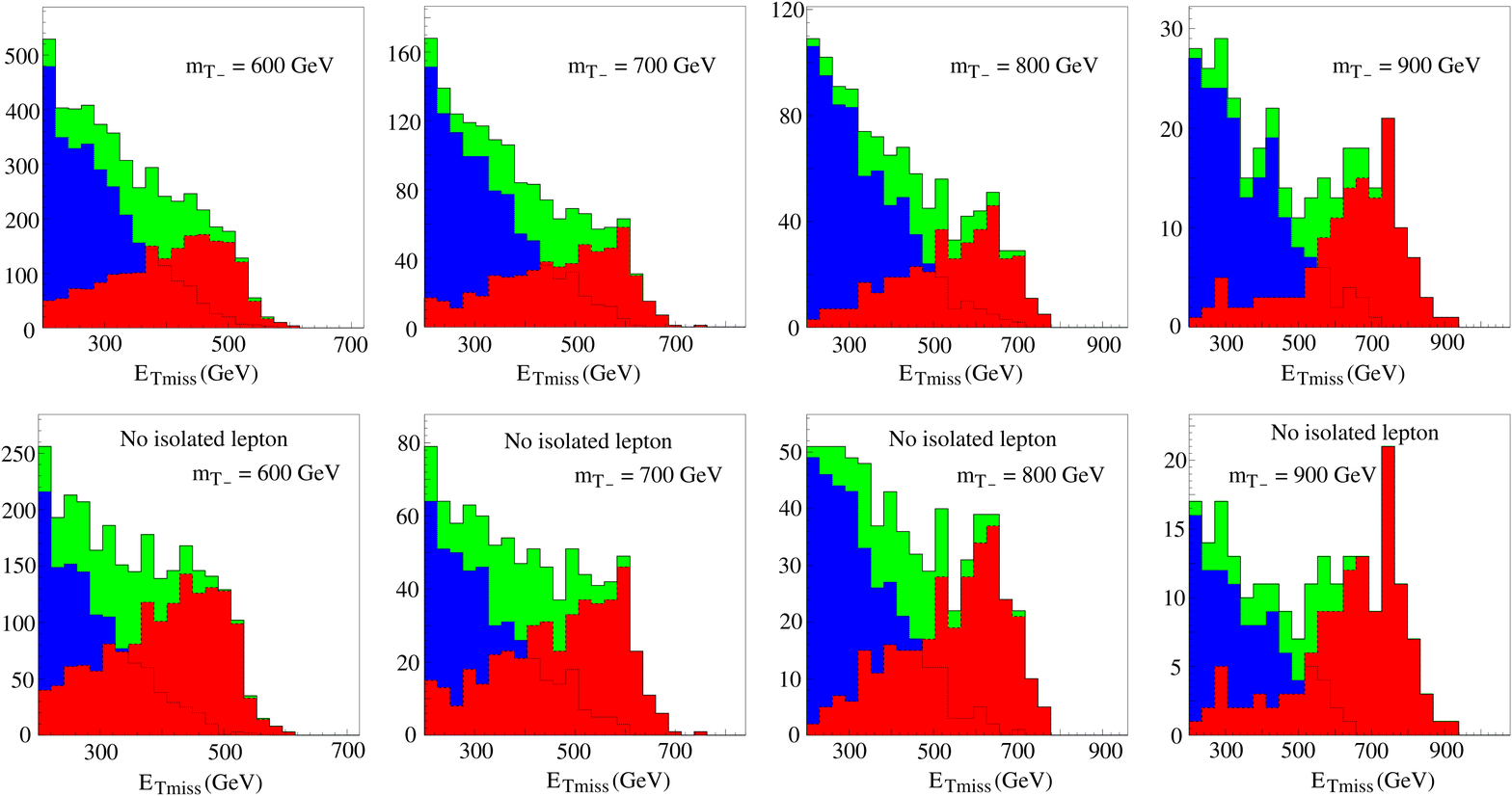}}
\caption{\small $E_{\rm Tmiss}$ distributions for points I to IV from left to right,
         with $2m_{T_-}$ $-$ 200 (300)~GeV $< M_{\rm eff} <$ $2m_{T_-}$ for $m_{T_-}$
         $\leq 700$ ($\geq$ 800)~GeV. The red and blue histograms are for the
         signal and background distributions, and the green histograms are the sum
         of these. $m_{\rm min}(j j j) <$ 200~GeV is required at least in one of
         the hemispheres. In the bottom four figures, only the events without
         isolated leptons have been used.}
\label{etmiss}
\end{center}
\end{figure}

Using the nature of these production and decay kinematics, we can find the
kinematical region with good separation between signal and background. We
restrict $M_{\rm eff}$ to a certain large value, so that we see the bump of the
signal event in the $E_{\rm Tmiss}$ distribution. In Fig.\ref{etmiss}, we show the
signal and background $E_{\rm Tmiss}$ distributions for points I to IV (top four
figures). Each plot corresponds to the integrated luminosity $\int d t {\cal
L} = 50$ fb$^{-1}$. We restrict $M_{\rm eff}$ to $2m_{T_-} -$ 200~GeV $< M_{\rm
eff} < 2m_{T_-}$ for $m_{T_-} =$ 600 and 700~GeV, and $2m_{T_-} -$~300 GeV $<
M_{\rm eff} < 2m_{T_-}$ for $m_{T_-} =$ 800 and 900~GeV, so that $E_{\rm Tmiss}$
becomes maximal and the signal rate is still reasonably high. Here the top
mass cut, $m_{\rm min }(j j j) < 200$~GeV, is required for at least one of the
hemispheres, which reduces the background by factor of 5 and the signal by
factor of 3 compared to the case where no cut is applied on $m_{\rm min}(j j j)$. In the
bottom four figures in Fig.\ref{etmiss}, we further require that the events
have no isolated leptons. The isolated lepton is produced by the leptonic decay
of the top quark. The $t\bar{t}$ background is reduced by a factor
of two by this cut, with no significant reduction of signal events. Each
distribution shows the clear excess of events over the (exponentially
decreasing) background.

\begin{figure}[p]
\begin{center}
\scalebox{0.2}{\includegraphics*{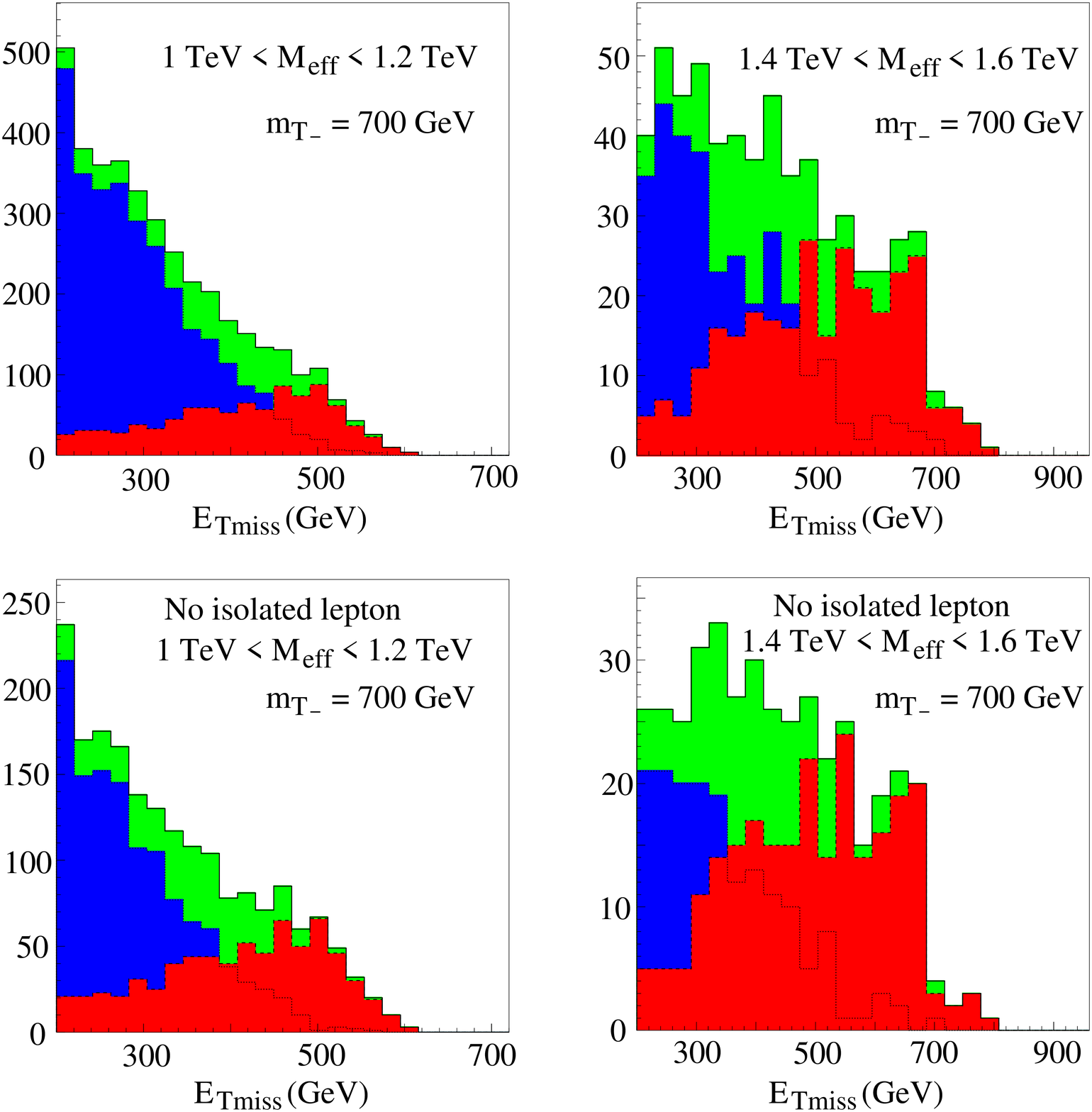}}
\caption{\small $E_{\rm Tmiss}$ distributions for point II with 1000~GeV
         $<M_{\rm eff}<$ 1200~GeV (left two figures) and 1400~GeV
         $<M_{\rm eff}<$ 1600~GeV (right two figures). $m_{\rm min}(j j j) <$ 
         200~GeV is required for at least one of the hemispheres.}
\label{etmiss1}
\end{center}
\end{figure}

The excess is less prominent for smaller $M_{\rm eff}$ bins. We show the
$E_{\rm Tmiss}$ distribution for point II with 1000~GeV $< M_{\rm eff} <$1200~GeV
in Fig.\ref{etmiss1} (left two figures). While the number of the signal
increases by factor of two, the $t\bar{t}$ background increases more than
factor of three. On the other hand, if one increases $M_{\rm eff}$, the number
of the signal reduces rather quickly, and we find that the events in this
region do not contribute much to the discovery of the $T_-$ quark. See
Fig.\ref{etmiss1} (right figure) for the distribution with 1400~GeV $<
M_{\rm eff} <$1600~GeV. Note that the events in the region $M_{\rm eff} \gg
2m_{T_{-}}$ arise from highly boosted $T_-$ quarks, therefore the fraction
$E_{\rm Tmiss}/M_{\rm eff}$ decreases, making the signal distribution less prominent
over the background.

\begin{table}[t]
\center{
\begin{tabular}{|c||c|c|c|c|c|}
\hline
  $m_{T_-}$               & $M_{\rm eff}^{\rm min}$ & $E_{\rm Tmiss}^{\rm cut}$
& Signal/BG              & Signal/BG             & Signal/BG           \\
  (GeV)                  & (GeV)                 & (GeV)
& (0-lepton with top cut) & (with top cut)         &                     \\
\hline
\hline
 600 & 1000& 400 & 842/106 & 1053/313 & 3336/1304 \\
\hline
 700 & 1200& 450 & 263/54  &  332/114 & 1284/582  \\
\hline
 800 & 1300& 500 & 208/28  &  249/57  &  874/417  \\
\hline
 900 & 1500& 550 &  93/7   &  105/16  &  397/203  \\
\hline
\end{tabular}
}
\caption{\small The signal to background ratio at the sample points I to IV. We
         take a region with $M_{\rm eff}^{\rm min} < M_{\rm eff} < 2m_{T_-}$ and
         $E_{\rm Tmiss}> E_{\rm Tmiss}^{\rm cut}$. The ratio is best if one requires the
         top cut and vetoes isolated leptons.}
\label{table:SN}
\end{table}

The numbers of the signal and background events in the signal region are shown
in Table \ref{table:SN}. Here we take the same signal region as that of
Fig.\ref{etmiss}. The signal to background ratio with the top cut is more than
3 for all sample points. Thus, it is clearly shown that the signal dominates
in the region where $E_{\rm Tmiss} \sim 0.5 M_{\rm eff}$.

\subsection{Statistical significance for the $T_-$ quark discovery}

As shown in Fig.\ref{etmiss}, the $t\bar{t}$ background
reduces rather quickly at large $E_{\rm Tmiss}$. 
Near the end of the distribution of $E_{\rm Tmiss}$, 
the distribution is dominated by the signal in Fig. \ref{etmiss}. 
On the other hand, the signal distribution decreases quickly after its peak. 
We therefore fit the decrease of  total distribution 
 near the maximum value $\sim 0.5 M_{\rm eff}$ to the Gaussian
function 
\begin{eqnarray}
 F(E_{\rm Tmiss})
 =
 h\exp\left(-\frac{(E_{\rm Tmiss} - E^{({\rm avg})}_{\rm Tmiss})^2}{2\sigma^2}\right)~.
\end{eqnarray}
The result is summarized in Table \ref{table:endfit}, where we have
used six or seven bins with the bin size ($\Delta E_{\rm bin}$) between
20.8 and 35.2~GeV. These fits give $\Delta\chi^2/(N_{\rm bin} - 3) \sim 1$.

The statistical significance of the signal is given by $h/\Delta h$, where $h$
is the height of the Gaussian distribution and $\Delta h$ is its error. The
significance is more than seven for all sample points. 
When we fit the background distribution to the same Gaussian function, 
we obtain $\sigma=\sigma_{t\bar{t}}$ as 102(125)~GeV for the fit 
above $E_{\rm Tmiss}>304(328)$~GeV for 
1000 GeV $< M_{\rm eff} <1200$~GeV (1200 GeV $< M_{\rm eff} <1400$~GeV)
respectively. 
The $\chi^2$ of the fit is   $\Delta \chi^2/(N_{\rm bin} - 3) =$ 1.1 (1.2). 
We find ¡¡$\sigma \ll \sigma_{t\bar{t}}$, therefore, the detected edge is clearly
inconsistent with the $t\bar{t}$ background distribution.


\begin{table}[t]
\center{
\begin{tabular}{|c||c|c|c|c|c|c|}
\hline 
  $m_{T_-}$ (GeV) & $h$ & $\Delta h$ & $\sigma$ (GeV) & $N_{\rm bin}$
& $\Delta {E_{\rm bin}}$ (GeV) & $h/\Delta h$ \\
\hline
\hline 
 600 & 190 & 14.7 & 38.5& 7 & 20.8 & 13 \\
\hline
 700 & 160 &  6.8 & 43  & 6 & 25.6 & 24 \\
\hline
 800 &  46 &  4.9 & 66  & 6 & 30.4 & 9.4 \\
\hline
 900 &  17 &  2.4 & 76  & 6 & 35.2 & 7.1 \\
\hline 
\end{tabular}
}
\caption{\small Fit of the $E_{\rm Tmiss}$ distribution to the Gaussian function
         near $0.5M_{\rm eff}$. Here, $N_{\rm bin}$ is the number of bins used
         for the fit and $\Delta E_{\rm bin}$ is the bin size.}
\label{table:endfit}
\end{table}



\begin{figure}[t]
\begin{center}
\scalebox{0.27}{\includegraphics*{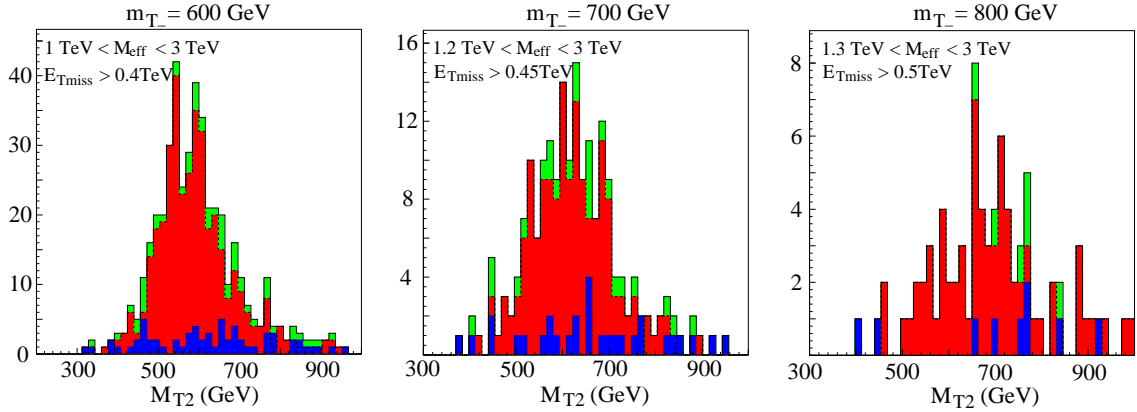}}
\caption{\small Distribution of the $M_{T2}$ variable for the events in the
         signal region defined in Table \ref{table:SN}. 
         $m_{\rm min}(j j j)<$ 200~GeV is
         required in at least one of the hemispheres. $m_{T_-} =$ 600, 700, 800~GeV from left to right.}
\label{fig:mt2}
\end{center}
\end{figure}

We found that the signal is prominent over the background when  $M_{\rm eff}$ is
restricted to the region near $2m_{T_{-}}$. We therefore calculate the $M_{T2}$ variable for the
events near the bump. We use ``Cambridge $M_{T2}$'', which is defined as
\begin{equation}
 M^2_{T2}
 =
 \min_{p_{1T}^{A_H } + p_{2T}^{A_H} =P_{\rm Tmiss}}
 \left[
  \max [m_{T_-}^2(p^{\rm vis}_1, p^{A_H}_1), m_{T_-}^2(p^{\rm vis}_2, p^{A_H}_2)]
 \right]~.
\end{equation}
It is a function of the transverse momenta and masses of the two visible particles,
$P_{\rm Tmiss}$, and the mass of the invisible particle. $M_{T2}$ is sensitive to the
mass difference $m_{T_-} - m_{A_H}$, but is not sensitive to the overall mass
scale. We take events where there are more than three jets with $p_T >$50~
GeV in each hemisphere. We calculate $M_{T2}$ by taking the visible particle 
momenta as  the sum of the three jet momenta used to calculate the minimum
three jet mass $m_{\rm min}(j j j)$ of the hemispheres, and we fix $m_{A_H} =$ 150~GeV. We
also require $m_{\rm min}(j j j) <$ 200~GeV for at least one of the hemispheres.
The distributions are shown in Fig.\ref{fig:mt2}. We found a positive
correlation to the mass of the top partner, however the number of events that
can be used for this analysis is small.

\subsection{Difference between $T_-$ and scalar top signals}

We discuss the differences in the production and decay distributions between the
$T_-$ quark and scalar top ($\tilde{t}$) signals. It is impossible to find the
process $pp \rightarrow \tilde{t}\tilde{t}^*$ followed by the decay
$\tilde{t}\rightarrow t \tilde{\chi}^0_1$ at the LHC, because the production
cross section is a factor 10 smaller than the $T_-$ quark
production cross section. However, we can still learn something from the
comparison.

Here we take $m_{T_-} = m_{\tilde{t}} =$ 800 GeV and $m_{A_H} =
m_{\tilde{\chi}^0_1} =$ 150~GeV. We took low energy parameters of the MSSM so
that all particles except stop and the LSP ($\tilde{\chi}^0_1$) are too heavy to be
accessible at the LHC, and $\tilde{t}$ decays into $t$ and $\tilde{\chi}^0_1$ 
with 100\% branching ratio. The events for the $pp \rightarrow
\tilde{t}\tilde{t}^* \rightarrow t\bar{t}\tilde{\chi}^0_1\tilde{\chi}^0_1$ 
process are
generated by the CompHep code and interfaced to HERWIG.

\begin{figure}[t]
\begin{center}
\scalebox{0.26}{\includegraphics*{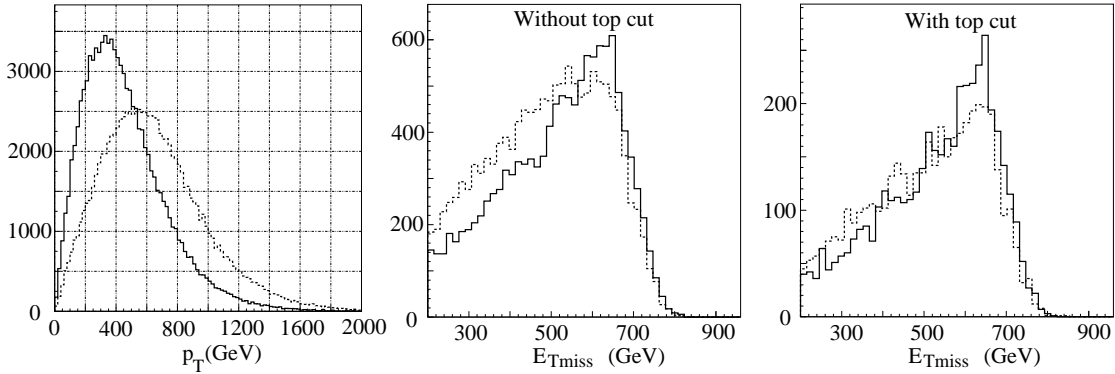}}
\caption{\small The $p_T$ distribution of $T_-$ (solid line) and $\tilde{t}$
         (dashed line) at the LHC (left figure). The $10^5$ events are
         generated by the CompHep code. We set $m_{T_-} = m_{\tilde{t}} =$ 800~GeV.
          The $E_{\rm Tmiss}$ distribution for $T_-$ (solid line) and $\tilde{t}$
         (dashed line) productions with 1300~GeV $< M_{\rm eff} <$ 1600~GeV
         (center and right figures). For the right figure, we required
         $m_{\rm min}(j j j) <$ 200 GeV for at least one of the hemispheres.}
\label{fig:comparison}
\end{center}
\end{figure}

In Figure \ref{fig:comparison} (left figure), we show the $p_T$ distributions of
$T_-$ and $\tilde{t}$ production at the LHC. The cross section is dominated by
the low $p_T$ component for the $T_-$ quark. On the other hand, the $\tilde{t}$
production cross section is dominated by P wave, therefore the $p_T$
distribution is broadened. The production cross sections peak at 350 and 600 GeV
for the $T_-$ quark and $\tilde{t}$, respectively,  where $\beta \sim$ 0.4 and
0.6. Both cross sections are kinematically suppressed near the peak by $\beta$.
Especially in the stop case, the squared amplitude is proportional to
$\beta^2$ due to the P wave production. The cross section beyond the peak is
suppressed by the quickly decreasing quark and gluon parton distribution
functions. If this $p_T$ distribution is reconstructed from the $t\bar{t}$
distribution, this indirectly suggests that the produced particle is a fermion.

The difference of the distributions affects the $E_{\rm Tmiss}$ distribution for 
fixed $M_{\rm eff}$ as shown in Fig.\ref{fig:comparison} (center and right 
figures). Here, the center figure shows the event distribution with 1300~GeV $<
M_{\rm eff} <$ 1600~GeV, and, in the right figure, when the top cut is required for
at least one of the hemispheres. The $E_{\rm Tmiss}$ distribution of the $T_-$
events has a sharper peak near $0.5M_{\rm eff}$, while the distribution is
broader for $\tilde{t}$. This is because $\tilde{t}$ is produced with higher
$p_T$, therefore the top quark has, on average, a larger energy. Therefore the
average fraction $E_{\rm Tmiss}/M_{\rm eff}$ becomes smaller.

Finally we comment on the top reconstruction for the supersymmetric signature. 
In supersymmetric models, the top quarks mostly arise from gluino decays through
processes, $\tilde{g} \rightarrow \tilde{t}t$, $\tilde{g} \rightarrow
\tilde{b}b \rightarrow t b\tilde{\chi}^+$. The top quark in the event is
hardly seen in the $m_{\rm min}(j j j)$ distribution due to the additional high
$p_T$ jets and leptons. In previous studies, the top quark is therefore
searched for by looking for the jets consistent with the top decay kinematics, namely
$m(j j) \sim$ 80~GeV and  $m(b j j) \sim$ 175~GeV by looking for the combination
of jets $i$, $j$, and $k$, which minimizes $\Delta\chi^2$ defined as
\begin{equation}
 \Delta\chi^2
 =
 \frac{(  m(i,j) - m_W)^2}{(\Delta m_W)^2}
 +
 \frac{(m(i,j,k) - m_t)^2}{(\Delta m_t)^2}~,
\end{equation}
where $\Delta m_W$ and $\Delta m_t$ are the expected errors on $W$ and top mass
reconstruction.

The efficiency to find the top quark in the events becomes higher if we look
for the jet combination consistent with top decay kinematics, however, in this
case, we occasionally find top candidates in events which do not contain
parton level top quarks, increasing backgrounds from the other SM processes
\cite{SUSYFilip}. The signal distributions may also be distorted by such a procedure.
This is particularly the case when the number of jets in the events is
large. A scheme to subtract the accidental background
from the top decay kinematical distributions has been developed \cite{Hisano:2003qu}.

\section{Discussion and outlook}
\label{Summary}

In this paper, we have studied the phenomenology of the top partner in the littlest
Higgs model with T-parity. This model predicts a stable neutral massive gauge
boson ($A_H$), and a relatively light top partner ($T_-$). The EW precision
measurements and dark matter observations constrain parameters of the
model, $f$, $m_h$, $\lambda_1$ and $\lambda_2$. We find $m_{A_H}$ and $m_h$ are
strongly constrained by the relic density constraint, if coannihilation
processes with other T-odd particles are not efficient. Combined with the EW
precision measurements, we have found the lower limit of $m_{T_-}$ as a
function of $f$, and clarified the fact that the top partner mass must be above
600 GeV and the mass difference between $T_-$ and $A_H$ is large. The signature
of $T_-$ quark pair production at the LHC is therefore high $p_T$ top
quarks from $T_- \rightarrow t A_H$ and missing $p_T$ coming from the $A_H$.
We have studied the signal and background distributions at the LHC.

Unbiased reconstruction of the top quark is essential to establish the
existence of the top partner without increasing accidental backgrounds. We
apply the hemisphere analysis, which is an algorithm to find two axes in
events originating from particles produced in pairs. We have found that the
appropriately defined three jet invariant mass in a hemisphere shows a clear peak
at the top quark mass with a tail due to mis-reconstructed events. The three jet
invariant mass is calculated from the pair of jets giving the maximum
invariant mass in a hemisphere and a jet giving the minimum three jet invariant
mass when combined with this jet pair. 
The efficiency to reconstruct at least one  top quark, i.e. a hemisphere with 
$m_{\rm min}(j j j) <$ 200~GeV in $T_{-}\bar{T}_{-}$ events, is reasonable
$\sim 20\%$.

The dominant background to the $T_-\bar{T}_-$ process is $t\bar{t}$
production, whose production cross section with $E_{\rm Tmiss}>$ 400~GeV is 
${\cal O}(10)$ or
more times larger than the signal cross section. We study the $E_{\rm Tmiss}$ distribution
as a function of $M_{\rm eff }$. Due to a simple kinematic reason, the
separation of the signal from background becomes best when $M_{\rm eff} \sim 
2m_{T_{-}}$. A pseudo-edge structure of the signal can be observed over the
$t\bar{t}$ background in that region, and the significance of the signature
turns out to be larger than $7\sigma$ for the case of $m_{T_-} \leq$ 900~GeV.

The reconstructed signal is clearly different from that of SUSY events. In the
case of SUSY, the production cross section of $\tilde{t}\tilde{t}^*$ is
rather small and is not detectable. Scalar top quark may be produced
from gluino decay. In that case, additional jets are also produced, and the top
cut $m_{\rm min}(j j j)<$ 200~GeV for the jets in a hemisphere can hardly be
satisfied.

Results of our simulation given in this paper should be regarded as an order of magnitude
estimate. We have used the simple smearing and jet finding algorithms provided by
the AcerDET code. The $E_{\rm Tmiss}$ and jet energy resolution could be different
at the ATLAS and the CMS detectors in the LHC environment. Note that the energy
resolution assumed by the AcerDET is closer to the ATLAS detector (
The default is $\Delta E/E=50\%/\sqrt{E}$, while $\Delta E/E
= 50 (100)\%/\sqrt{E}$ for the ATLAS (CMS) detectors, respectively).
Furthermore, the background from $t\bar{t}$ production is estimated by the
HERWIG code in this paper, where $\sigma(t\bar{t}) = $ 400~pb. However, it is
well known that the $t\bar{t}$ cross section receives large NLO corrections of
around a factor of two. The multi-jet final state $t\bar{t}$ + jets should also
give an important background to the events with $E_{\rm Tmiss}$. For the parameters
we have studied, S/N is high in the signal region. Moreover, the signal has a
peak structure in the $E_{\rm Tmiss}$ distribution. Therefore, we believe that our
result is stable against additional sources of background and a
factor of two
increase of the background. In addition, $t\bar{t}$ production with
a multi-jet final state has a better chance to have a large three jet invariant
mass in a hemisphere, because events have additional jets which are not
kinematically constrained by the top mass shell conditions. It is likely that
the process $t\bar{t}$ + jets would be rejected by the top cut
$m_{\rm min}(j j j) <$ 200~GeV. It would be interesting to study the processes
$t\bar{t}$ + jets and $T_{-}\bar{T}_{-}$ + jets to check this expectation
explicitly.

In this paper, we do not study mass reconstruction in detail. The mass
difference $m_{T_-} - m_{A_H}$ could be extracted from the $M_{\rm eff} -
E_{\rm Tmiss}$ distribution if the background distribution can be calibrated
precisely. We have also looked into the $M_{T2}$ distribution for the sample
with at least one reconstructed top quark, which shows positive correlation
with the mass difference. The fit may be improved if the efficiency to
reconstruct the top quark is improved by selecting the jets consistent with the
top decay kinematics. The 
combinatorial
backgrounds may be reduced if we select
the top candidate among the jets in a hemisphere.

The existence of the top partner and dark matter associated with the
T-parity symmetry is an important feature of physics beyond the Standard Model
as we discussed in the introduction. The collider phenomenology of the top
partner therefore deserves more realistic studies. We hope more realistic
studies will be performed by the ATLAS and the CMS groups in the future.

\vspace{1.0cm}
\hspace{0.2cm} {\bf Acknowledgments}
\vspace{0.5cm}

This work is supported in part by the Grant-in-Aid for Science Research, 
Ministry of Education, Culture, Sports, Science and Technology, 
Japan (No.16081211 for S.M. and
No.16081207, 18340060 for M.M.N.).

\end{document}